\documentclass{scrartcl}
\usepackage[utf8]{inputenc}
\usepackage[T1]{fontenc}
\usepackage[british]{babel}
\usepackage{graphicx}
\usepackage{amsmath}
\usepackage{libertinus}
\usepackage{xurl} 
\usepackage[hidelinks]{hyperref}
\usepackage{changepage}
\usepackage{diagbox}
\usepackage{xcolor}
\usepackage{wasysym}
\usepackage{tikzsymbols}
\usepackage[round]{natbib}
\setcitestyle{aysep={}}
\usepackage{enumitem}
\ifdefined\LaTeXML
  \author{Fredrik Jansson\\[0.6em]
    \textit{Department of Business and Mathematics, M\"alardalen University, V\"aster\aa s, Sweden}\\
    \textit{Centre for Cultural Evolution, Department of Psychology, Stockholm University, Sweden}\\
    \textit{Institute for Futures Studies, Stockholm, Sweden}}
  \date{Preprint\footnote{Chapter in Johan Lind and Anna Jon-And, eds. (in press): \emph{Cultural Evolution from Minimal Principles}. Cambridge, United Kingdom: Cambridge University Press.}\footnote{This work was supported by the Marianne and Marcus Wallenberg Foundation (2021.0039).}}

\else
  \usepackage{authblk}
  \author{Fredrik Jansson}
  \affil{\vspace{1em}Department of Business and Mathematics, M\"alardalen University, V\"aster\aa s, Sweden}
  \affil{Centre for Cultural Evolution, Department of Psychology, Stockholm University, Sweden}
  \affil{Institute for Futures Studies, Stockholm, Sweden}
  \date{Preprint\footnote{Chapter in Johan Lind and Anna Jon-And, eds. (in press): \emph{Cultural Evolution from Minimal Principles}. Cambridge, United Kingdom: Cambridge University Press.} \footnote{This work was supported by the Marianne and Marcus Wallenberg Foundation (2021.0039).}\\[-1em]}
\fi

\AtBeginEnvironment{thebibliography}{\raggedright}
\setlist[itemize]{noitemsep}
\setlist[enumerate]{noitemsep}
\ifdefined\LaTeXML
  \providecommand{\setcapindent}[1]{}
  \providecommand{\setkomafont}[2]{}
  \providecommand{\addtokomafont}[2]{}
\fi
\setcapindent{0pt}
\setkomafont{captionlabel}{\bfseries}

\title{The dynamics of cultural systems}

\begin{document}

\maketitle

\begin{abstract}
    \noindent Culture is not just traits but a dynamic system of interdependent beliefs, practices and artefacts embedded in cognitive, social and material structures. Culture evolves as these entities interact, generating path dependence, attractor states and tension, with long-term stability punctuated by rapid systemic transformations. Cultural learning and creativity is modelled as coherence-seeking information processing: individuals filter, transform and recombine input in light of prior acquisitions and dissonance reduction, thereby creating increasingly structured worldviews. Higher-order traits such as goals, skills, norms and cognitive gadgets act as emergent metafilters that regulate subsequent selection by defining what counts as coherent. Together, these filtering processes self-organise into epistemic niches, echo chambers, polarised groups and institutions that channel information flows and constrain future evolution. In this view, LLMs and recommender algorithms are products of cultural embeddings that now act back on cultural systems by automated filtering and recombination of information, reshaping future dynamics of cultural systems.

\end{abstract}

\section{Introduction}

You are walking through a city square and notice someone wearing jeans with large rips across the knees and thighs. For a moment, you wonder: Did they fall? Are they poor? After seeing even more people with ripped jeans, you start to wonder whether there is some practical reason for those tears. However, soon you also notice a pattern: the wearers are mostly young, fashion conscious, and stylistically coordinated. The rips are clearly no accident, nor do they seem to fulfil a practical function, but rather, they are part of a look.

At that moment, the jeans shift in your perception, from damaged clothing to a social signal. But what they are signalling depends entirely on context. In one setting, they may signal casual trendiness; in another, they echo punk rebellion or grunge nonconformity. A few decades earlier, they might have evoked poverty or working-class toughness. The same physical trait, ripped fabric, carries wildly different meanings, depending on the web of cultural associations it plugs into.

Ripped jeans are not just a trait, but a node in a cultural system. They co-evolve with other traits that reinforce or redefine their meaning. In the punk era, they made sense within a web of anti-establishment symbols. In fast fashion, they became a curated aesthetic. Cultural change here is driven not by the jeans themselves, but by how they fit into shifting systems of meaning. The way they fit into the environment and the reactions they evoke are also very different between a rock concert and the Nobel Banquet.

Cultural evolution here is driven by reconfigurations in how traits relate to each other. These clusters of traits evolve not as isolated elements, but as interdependent packages. To understand their spread or transformation, we may need to examine how traits relate to one another, not just their individual appeal. Imagine attending a communion and trying to make sense of why people kneel to a man in a white robe while he feeds them wine and wafers and talks about flesh and blood, without being aware of the context.

Not everything is moving parts with arbitrary representations, though. Wine and blood do have about the same colour and ripped jeans are not that impractical. Agricultural practices reflect seasonal cycles. Scientific theories consist of interrelated definitions, hypotheses and processes, and they survive based on internal consistency and ability to make predictions about the world. Tools and artefacts are constrained by biomechanics and physics, and we see technological development of increasingly capable tools because artefacts are combined in ways that satisfy these constraints, and certain combinations are selected over others by virtue of their fit with our needs and cultural practices.

There is an interplay between technology and norms: for example, driving on the right-hand side of the road correlates with steering wheel placement around the world. However, until 1967, Sweden was an anomaly in having left-hand traffic and left-side steering, because cars were mostly imported from continental Europe and the US. This led to reduced visibility for overtaking and made driving less safe. This could be seen as a systemic tension that facilitated and may have partly driven the transition to right-side driving.

It seems that the interplay between cultural traits is a key driver of cultural evolution; we will indeed see more examples of this in this chapter. Not surprisingly, then, ideas about the structural properties of culture are central to anthropology, semiotics, sociology, linguistics, and science and technology studies. Early “proto-structuralist” perspectives \citep{durkheim1912elementary, mauss1954gift} viewed cultural forms as parts of an integrated whole, sustained by shared representations and collective practices. Structuralism \citep{levi-strauss1963structural, desaussure1959course} developed this into an analysis of culture as systems of interdependent elements, whose meanings arise from their relations. Poststructuralist approaches \citep{foucault1972archaeology, foucault1977discipline, bourdieu1977outline, bourdieu1990logic} added attention to historical contingency, generative processes, and the reconfiguration of relations in ways that carry an implicit evolutionary element, though without an explicit theory of cumulative change. In the philosophy of science, the paradigm model \citep{kuhn1962structure} similarly framed scientific knowledge as structured by shared conceptual schemes that shift through systemic transformations. Parallel systems perspectives in sociology \citep{luhmann1995social} and science and technology studies \citep{latour2005reassembling} likewise treat change as reconfigurations in the web of associations linking traits, practices, and artefacts. Bringing an explicit cultural evolutionary systems perspective into this picture not only integrates these insights but also addresses several persistent challenges in cultural evolution theory itself.

In another chapter \citep{jansson2026cultural}, we made an orientation of the different schools of cultural evolution along with some of their challenges. In memetics, cultural evolution is a Darwinian process based on memes, which are replicators in their own right, residing in human brains \citep{dawkins1976selfish, dennett1995darwins, blackmore1999meme}. An outstanding issue is to determine what a meme is, that is, the unit of imitation, and how memes interact. One proposal is that memes often cluster into larger constellations, or memeplexes, whose elements mutually reinforce each other’s retention and transmission, such as religious doctrines, political ideologies, or scientific paradigms \citep{blackmore1999meme, aunger2000darwinizing}. From a cultural-systems perspective, memeplexes may be treated as emergent wholes in which individual memes form interdependent components, connected through both functional roles and symbolic associations. Such systems can be hierarchically organised, with stability arising from patterned relations among parts that may themselves be composite \citep{buskell2019systems}. This makes the “unit of culture” a matter of analytical granularity, with memes nested within memes all the way down.

Cultural attraction theory \citep{sperber1996explaining, miton2023cultural} focuses on biased transformation rather than selection \citep{elmouden2014cultural, mesoudi2021cultural}. There is an epidemiology of representations rather than a direct spread of cultural traits. People’s representations are often similar because they reconstruct information towards cultural attractors. What are then these attractors: what constitutes them and where do they come from? This may again be a matter of granularity \citep{acerbi2015if} and trait interactions. Some traits may form clusters that reinforce one another, and adopting or triggering one of those traits may lead on to a path towards adopting or activating the other traits of the same cluster \citep{buskell2019systems, miton2022cultural}.

In what has been referred to as the ``standard model'' \citep{boyd1985culture, sterelny2017cultural}, like in memetics, there is a key focus on selection. In contrast to memetics, the theory also addresses the origin of culture and human abilities. One way to explain the stability of culture and recurrent patterns is that genes and culture may have co-evolved to provide humans with social learning mechanisms that are selective with different biases \citep{boyd1985culture, henrich2016secret}, like prestige \citep{henrich2001evolution} and conformity \citep{henrich2001cultural}. However, empirical evidence from experimental and developmental psychology suggests that such strategies may emerge through social interaction and associative learning, especially during childhood \citep{heyes2018cognitive}. In this view, many social learning strategies are not innate special-purpose adaptations but cognitive gadgets: culturally constructed, domain-general abilities assembled through experience and interaction. Through trait interactions, previously acquired traits will bias individuals' further acquisition of traits, and thus allow biases to be culturally emergent. There is not only a co-evolution between genes and culture, but between culture and other culture. Systems thinking may prove cultural co-evolution to be highly potent.

\section{Embeddings}\label{sec:embeddings}

It should be clear by now that cultural traits, such as beliefs, symbols, practices and artefacts, are not free-floating entities but are embedded in broader cognitive, social, and material systems that shape their meaning, function and persistence. Cognitive embeddings situate cultural information within mental models, schemas, and inferential frameworks that guide interpretation and integration \citep[e.g.][]{bartlett1932remembering, morin2015how, heyes2018cognitive}. Social embeddings position cultural traits within networks, institutions, and power structures that influence their reproduction and transformation \citep[e.g.][]{granovetter1985economic, white2008identity}. Material embeddings anchor culture in artefacts, infrastructures, and ecological settings that serve both as external symbolic storage \citep[e.g.][]{donald1991origins, sterelny2012evolved} and as physical or technological constraints on possible variation \citep{basalla1988evolution, arthur2009nature}. Because of such embeddings, cultural change is typically constrained and path dependent, with shifts in one element often cascading through the interconnected system.

\subsection{Trait relations}

So, where do the relations between traits come from? Ripped jeans and religious ceremonies are embedded in rich webs of relations that are culturally emergent. There are numerous other examples, such as beliefs that tend to co-occur \citep{boutyline2017belief}, forming ideological packages \citep{kahan2016politically}, where one belief tends to come with often seemingly unrelated beliefs, for example believing in anthropogenic climate change and to favour gun regulation in the US, or opposing same-sex marriage while supporting military spending \citep{boutyline2017belief}. Ideologies are in turn associated with lifestyles, with latte liberals as a prominent example \citep{kahan2016politically}. These associations are so frequent and encoded in our language that the success of LLMs partly rests on their ability to capture such systemic properties of culture, reproducing the patterned co-occurrences that make cultural communication coherent.

These cultural associations are seemingly unrelated, though are they completely arbitrary? They may seem so, studied as small subsystems, but may make more sense when including confounding traits and historical trajectories. There may be evolutionary stories explaining their co-occurrence provided relations between other traits. Both latte and liberal signal urban, educated, cosmopolitan lifestyles, often tied to higher income and openness to foreign influences. Attitudes towards abortion rights and stem cell research tend to be mediated by religious beliefs \citep{nisbet2005competition}. If we dig deep enough, we might eventually find material associations and constraints; not necessarily that such constraints will lead to easily predictable, unequivocable cultural associations further down the line, but they may restrain the evolutionary pathways and then fixate them through path dependence. The alignment between support for climate policies and left-wing politics might illustrate this dynamic. In the 1970s and 1980s, even conservative leaders such as Margaret Thatcher voiced strong concern over climate change \citep{kenny2024changing}, but the economic constraints of ambitious climate action, such as regulation, subsidies, and carbon pricing, might have fit more readily with interventionist left-wing platforms. As progressive leaders profiled themselves on the issue and fossil fuel interests aligned with the right, partisan associations may have stabilised through path dependence \citep{mccright2011politicization}.

There are many examples of associations more directly subject to material constraints. Technology is one domain where evolution is driven by functional compatibilities between traits \citep{arthur2009nature}. Hydrogen passenger airships, amphicars, wearable jetpacks or three-wheeled ATVs (or for that matter cars) never took off because they were too impractical or unsafe, not to mention nuclear-powered vacuum cleaners, X-ray shoe-fitter machines and smell-o-vision films -- all great ideas, of course, but with not-so-compatible components. The incompatibility of the  steering wheel placement of Swedish cars with traffic rules was solved by replacing the rules instead.

Also more abstract domains, like beliefs, can be constrained by consistency. The scientific method is a practice that has proven highly useful and successful by enforcing such constraints \citep{popper1959logic, chalmers2013what}. This provides both a gradual development towards increasingly precise predictions and punctuated shifts as, over time even well-established para\-digms can accumulate anomalies, patched with ad-hoc assumptions until the inconsistencies become too great; in such moments of crisis, they are replaced by new paradigms that may reconfigure the very standards of evidence and explanation \citep{kuhn1962structure}.

Musical harmony may be a particularly clear example of interrelations between traits, with both material and cultural constraints. Individual tones have some intrinsic properties as to how they fit the range of the singer or instrument, but they make music only in combination with other tones. In these combinations, relations matter more than elements. The major triad is widespread not defined by the absolute pitches involved, but because of the relational pattern between root, third and fifth. From such relations emerge higher-order structures such as scales, harmonic progressions and styles, where each element gains significance only through its role in the whole. Different musical traditions organise these relations differently, but the systemic principle is the same: meaning arises from the configuration of parts, not from the parts themselves. As for material constraints, the major triad corresponds to low-order harmonic ratios in the natural overtone series, meaning that we are exposed to them every time we hear a voice, a violin or even a struck object. This property may render the cluster of tones with the relations of being a major third, a minor third and a fifth apart a highly stable one \citep{jan2022music}. Yet, the perfect ratios from the harmonic series have been mostly outcompeted in Western music by equal temperament, which only approximates the ratios so that the distance ratio is the same between any adjacent pair of notes. This related relation between tones has the benefit of enabling modulation and standardisation across instruments, and thus works better within its cultural ecosystem.

\subsection{Embeddings and co-evolution}

These examples show that not only is culture embedded in cultural ecologies, but also that relations between traits are drivers of cultural evolution. All examples included constraints, which apart from physical bounds are also found in cultural traits: especially norms and taboos regulate what traits are allowed in specified contexts. There are, however, also other types of co-dependencies.

Some traits are prerequisites for others. This is particularly evident in cumulative domains, such as technology. There would be no carts without wheels and no use of wheels without axles; writing on scrolls presumes a writing system, ink, some kind of pen and a writing surface; and writing this chapter on a computer presumes a tremendous amount of cultural artefacts and preexisting ideas. Innovations such as the printing press facilitated the spread of bible translations, which spurred the reformation \citep{eisenstein1980printing}; cheap contraception changed gender norms and family structure; and smartphones and internet connectivity is reshaping information pathways, whom we interact with and how we spend our time.

Other traits stabilise or preserve culture. Writing systems standardise and stabilise pronunciation \citep{labov1994principles} and content, such as stories, which are transmitted and recorded more faithfully. In oral traditions, rhythmic meter has a similar effect on stories. Institutional scaffolding such as schools, journals and courts actively retain certain traits by creating and enforcing rules, norms and infrastructures that filter, amplify and authorise certain variants over others. While stabilising the prevalence of other traits, these traits can give rise to strong path dependence and lock-in effects, from harmlessly strange symbols (the save function is still represented by a floppy disk) to inefficient standards. Rail gauges are different in neighbouring countries, hindering international connections, Betamax was said to be a technically superior video cassette to VHS, and there have been attempts at replacing the QWERTY keyboard by more efficient designs \citep{kay2013rerun}, but once established, these technologies built their own ecosystems (involving trains, VCRs and touch typing) that made them hard to replace. Minitel, a French predecessor to the internet, may have delayed the spread of internet connections in France \citep{mailland2017minitel}, and the imperial system of measurement has for some reason prevented the US from adopting the metric. Among the consequences of misalignment between two coexisting systems of measurement is a crash of the Mars Climate Orbiter in 1999, when NASA’s navigation team expected metric units while the manufacturer seems to have found US customary units more convenient.

Related to lock-ins and tight ecosystems are traits that, due to their high connectedness, cause large repercussions when altered. The earlier examples of innovations such as the printing press or smartphones had far-reaching cascading effects, but modifying existing traits can also incur vast cascades. Placing sun at the centre of the solar system instead of earth undermined Aristotelian cosmology, church authority and the human-centred worldview \citep{kuhn1957copernican}. Providing access to vernacular bibles and changing doctrinal authority in church paved the way for the reformation with vast effects on religious authority, literacy, state power, art, economy and social order \citep{macculloch2003reformation}. Linguistic traits can be so interdependent that a change in one element cascades through the whole system. The Great Vowel Shift in Middle English illustrates this: when one long vowel shifted upwards in tongue position, it encroached on the space of another vowel and left another space uninhabited, which impacted the neighbouring vowels in turn, triggering a domino effect across the entire vowel system. The result was a wholesale reconfiguration of English vowels rather than an isolated sound change (partly resulting in the unconventional mapping between spelling and pronunciation in Modern English) \citep{aitchison2001language}.

Co-dependencies can lead to co-evolution. Many religious beliefs and rituals are inseparable from the institutions that sustain them; religion both legitimises and is reinforced by collective institutions \citep{durkheim1912elementary}. Bureaucracy (and the organisation of society) is suggested to have co-evolved with writing \citep{goody1986logic}, and writing with literacy, which is both facilitated by and comes with a need for schools. Cars made dispersed suburban living practical and reorganised how cities are built \citep{jackson1985crabgrass}, and suburban living increases dependence on cars, again providing a lock-in that is at par with ideas of vibrant city centres and the green transition. To take some everyday examples, there are traits that can evolve on their own, but facilitate each other: knives and forks, coffee and sugar, shoelaces and bows, trousers and belts, smartphones and social media. Some traits evolve even more tightly, almost as packages, even though they can be used on their own: toothbrushes and toothpaste, shoes and socks, paper and pens, earphones and portable music players. Finally, many traits are almost co-constitutive: reading and writing, chess pieces and chess rules, musical scales and tuning systems, recipes and standardised measures, barcodes and checkout scanners.

Zooming out, we can see that culture, at almost any conceptual level, is composed of systems of interconnected components. These components can fit together to various degrees. When they fit well, we might expect the composition to be stable, with the different parts supporting each other, be it chords, technologies or religious institutions. Fitting many components however comes with increasing combinatorial challenges. Eventually, cultural systems can become unstable due to structural tensions. We have already seen some simple examples of this: the Swedish mismatch between the side of driving and the steering wheel placement, musical scales completely based on the harmonic series and the need to modulate for different instruments, and doctrinal authority in church and access to vernacular bibles.

A more drastic example is the increasing complexity and empirical tension of the Ptolemaic model of the “heavens”, with earth at the centre. To keep pace with new observations (non-uniform planetary speeds, retrograde loops, and changing brightness), geocentric astronomy layered auxiliaries onto the core model: off-centre circles, epicycles on epicycles, and Ptolemy’s equant (uniform motion about a point displaced from the centre). With ever more parameters, this belief system’s internal strain grew when trying to fit all components together while also making accurate predictions, yielding an increasingly baroque system tuned table-by-table to fit the sky \citep{kuhn1957copernican}. In Kuhn’s terms, this is normal science: committed puzzle-solving within a paradigm by adjusting instruments, procedures, and auxiliary hypotheses. As anomalies accumulate and fixes proliferate, coherence and simplicity degrade and a crisis threatens \citep{kuhn1962structure}. A Popperian view on science increases the tension, as accumulating ad hoc assumptions diminishes testability \citep{popper1959logic}.

As tension builds up, such a system comes to a point where patching is unviable and where either new traits may have cascading effects that lead to conceptual shifts, or the system is outcompeted by a more coherent one. Kuhn described such ruptures in science as paradigm shifts, moments when the existing framework can no longer sustain the accumulated anomalies and a new system of concepts, exemplars, and methods displaces the old. These shifts are not incremental but revolutions in worldview, where the same phenomena are seen through different lenses, with planets no longer circling Earth but orbiting the Sun, and diseases reframed from miasmas to microbes. This also fits Popper’s view of scientific progress: when a system becomes overburdened with ad hoc fixes, a more falsifiable and testable theory should replace it.

This process is not restricted to science. Sweden did not see a gradual shift towards more and more cars having right-side steering nor did more and more drivers shift to driving to the right (anyone trying would initially put great stress on the system and the person would soon stop being a cultural role model, one way or the other). Instead, the whole system was replaced at a specified time and day. Equal temperament in Western music illustrates a similar story. Successive tuning systems patched the tensions between pure intervals and the desire to modulate, until the accumulated fixes became unwieldy. The adoption of equal temperament replaced this patchwork with a coherent, if imperfect, system \citep{barbour1951tuning}. Other examples where rapid dramatic shifts might be ascribed to accumulating systemic tension and collapse are when different standards need to be combined, such as rail gauges, local time zones, charging cables and societal values. The last suggestion may need a few more elaborate examples:

For centuries slavery was upheld by legal, religious, and economic supports as stabilising components, but Enlightenment ideals, abolitionist mobilisation, slave revolts and shifts in the global economy created growing contradictions. Attempts at gradual reform or regulation proved unsustainable, and the system eventually collapsed throughout the western world in decisive breaks in the late 18th and 19th centuries \citep{drescher2009abolition}. The East German regime held together through censorship, surveillance, and Soviet backing, but economic stagnation, growing information flows, and reformist signals from Moscow undermined its coherence. Mounting protests and the opening of neighbouring borders created tensions that could not be contained, until a single press conference misstatement triggered a sudden systemic collapse on 9 November 1989 \citep{sarotte2014collapse}.

The examples are plentiful of how culture is embedded in a system of interconnected parts and how cultural evolution can be driven by how these elements reorganise in response to previous reorganisation, adapting to new conditions, and so on, leading to emergent patterns with disruptive shifts, self-organised by the elements of the cultural system. We can thus consider culture a complex adaptive system \citep[see][]{michaud2026complex}.

\subsection{Interaction with other systems}

Cultural systems are of course not enclosed vessels, but also interact with genetic preconditions and the physical environment. These provide the material constraints discussed earlier, but it should be stressed that the influence is bidirectional. Wherever humans go, we change the local environment by building houses, clearing land, using fire, irrigating fields, and domesticating plants and animals. Such practices vary with local conditions, from stilt houses in wetlands to stone shelters in highlands, illustrating how culture adapts to ecological pressures. At the same time, these practices transform environments in ways that alter the future conditions of life, both for humans and other species \citep[a process known as niche construction][]{odling-smee2003niche}. Cultural practices have transformed most of the land on our planet, changed the composition of life and its conditions in the oceans, and increased the degree of carbon dioxide in the atmosphere to levels that have led to global warming and significant climate change.

Genes and culture also form interacting systems \citep[known as gene--culture co-evolution;][]{boyd1985culture}. Genetic predispositions provide the prerequisites for cultural evolution and provide biases for particular forms of learning or perception, shaping the pathways of cultural evolution. Cultural practices in turn transform the ecological and social environments in which genetic selection occurs. A classic example is lactase persistence: the cultural system of cattle domestication and dairying created new nutritional niches, which in turn selected for genetic variants enabling adults to digest lactose \citep{itan2009origins, malmstrom2010high}.

Human behaviour and its evolution is thus a product of three co-evolving domains: biology, culture and the physical environment. They are successive levels of organisation in the history of complexity. The first major transition was the emergence of life from the physical and chemical processes of the early Earth, introducing genetic inheritance as a new mechanism for retaining and transmitting information. Later transitions in biological evolution, such as the emergence of chromosomes, eukaryotes, and multicellular organisms, illustrate how new levels of organisation arise from the integration of previously independent units \citep{maynardsmith1995major}. Human culture represents another such transition, where socially learned symbols and practices supplement and interact with genetic systems. Some scholars frame all of this under the rubric of cosmic evolution, tracing the emergence of complexity from physical laws, to life, to human societies \citep{chaisson2012singular, chaisson2001cosmic, spier2010big}. Each transition has brought not only new mechanisms of inheritance, but also increased speed and flexibility of adaptation: biological evolution unfolds faster than geological change, while cultural evolution can reconfigure systems within a single generation.

For analytic purposes, we may thus be able to treat the low‐level domains as slow constraints on higher ones. Over the time scales at which culture typically changes (seconds to generations), physical laws and the human genetic endowment are effectively constant, so cultural evolution proceeds within a relatively fixed biological capacity, and the evolution of both culture and life under invariant physical regularities. While feedbacks become visible over long horizons, cultural evolution in the space of lifetimes might be primarily understood by studying cultural systems, given the constraints imposed by the other systems.

However, within the confines of physical and biological systems, there are sometimes rapid changes that can impose external shocks on cultural systems, causing perturbations with far-reaching consequences. The Black Death of the 14th century, for example, decimated populations across Europe and thereby destabilised the feudal system, shifted economic power toward labourers and increased social mobility, leading to long-term shifts in societal organisation (\citealp{benedictow2004black}; perhaps even paving the way for universities and the spread of Christianity and nationalism, \citealp{herlihy1997black}). Similarly, the introduction of Old World diseases to the Americas during the Spanish conquest led to demographic collapse, the disintegration of indigenous polities and collapse of social structure, paving the way for European colonisation \citep{diamond1997guns}. Pandemics, famines, volcanic eruptions, oil shocks and nuclear disasters are other game changing events. Recently, the covid pandemic created an urgent need for remote work and education infrastructures, reshaping work cultures, schooling and mobility patterns in ways that remain in present-day cultural systems, even when social distancing is no longer needed (or even desired). Such shocks illustrate how cultural systems, though often appearing stable, can be rapidly redirected when external pressures overwhelm existing structures, leading to cascades of institutional, social, and symbolic change. In light of this, arms races, climate change and AI may soon come to put great stress on current cultural systems, with potentially revolutionary consequences.

\section{Information processing}

Cultural information is encoded partly in artefacts and infrastructures, not only in human language, such as texts and recordings, but also for example in practical tools, art and technology \citep{hutchins1996cognition, clark1998extended}. We have already seen numerous examples of systemic constraints on how different artefacts fit together, and natural selection will for example disfavour hull--keel--rig configurations on a boat that fail to maintain a positive righting moment, so the vessel capsizes easily \citep[cf.][]{rogers2008natural}. However, cultural information is also encoded symbolically in human brains, as knowledge, ideas, concepts, preferences and beliefs. This is mainly where we would expect cultural systems to evolve, since ideas can be rapidly added, lost and recombined \citep{boyd1985culture}. Our minds exercise a strong selection on cultural traits to build cohesive belief systems \citep{buskell2019systems, jansson2023cultural, jansson2021modelling}.

There is continuous feedback between macrolevel cultural systems and individual belief systems \citep[cf.][]{coleman1973mathematics}. Taken together, cognition is distributed across brains, tools, and social structures, and cultural evolution operates over this coupled system. Norms and institutions exert guidance and constraints for individual beliefs and actions and design information pathways between individuals through things like social hierarchies, urban planning, media, telecommunication etc. These entities provide individuals with information that needs to be accommodated in their belief systems and influence with whom to interact, what signals are legible and what options are thinkable. Individuals further reconstruct their belief systems through social transmission. Beliefs translate into behaviours (such as telling others about your beliefs), which are in turn observed by other individuals, potentially influencing their belief systems. Aggregated across networks, this reproduces or transforms macrolevel cultural systems, such as societal norms, institutions and technology. In this procedure, apart from forces like natural selection, information processing and proximal selection and transformation of cultural variants typically occurs in individual minds \citep[e.g.][]{boyd1985culture, sperber1996explaining, mesoudi2011cultural}. To understand the evolution of cultural systems, this would suggest that we need to understand the dynamics of belief systems, where beliefs are understood broadly as information entities living in our minds, including ideas and preferences.

\subsection{Preservative selection versus biased transformation}

The way beliefs are transmitted and acquired socially has been a matter of considerable debate. While the “standard model” \citep[e.g.][]{boyd1985culture} views this as a selection-like process, cultural attraction theory \citep[e.g.][]{sperber1996explaining} stresses the fact that traits are not just passed on from individual to individual, but information is reconstructed: the sender may mean one thing but their message may just trigger associations to something else in the receiver. If I say “red” then we might come to think of different nuances, and if I say "red is the colour of love”, then you might think I am doing a political statement, for example about socialism or American republicans, depending on your belief system. There is experimental evidence of both preservative selection-like transmission and biased transformation, for example in the transmission of urban legends \citep{eriksson2014corpses} and which process is most prominent depends on the type of information that is transmitted (\citealp{acerbi2015if}; see \citealp{jansson2026cultural} for a further discussion).

A systems approach might be able to capture both preservative and transformative processes in cultural evolution, phrasing them as system reconfigurations that look different depending on the level of granularity. We may first need to see social transmission as an event of cultural systems interacting rather than a single trait being picked from a sender’s repertoire and added to a container of traits representing the receiver’s repertoire.

If the sender is retelling the story of Little Red Riding Hood, then regarding this story as one trait, what ends up in the receiver’s mind is not an exact replica, but a transformed version of the story. The story is far from an atomic unit, however, and is built from many components that form a narrative together. Often, such a transmission event ends up with the receiver retaining a subset of the components told by the sender; ABCDEF turned into, say, ADE. The transformation of the story is a selection of its components. Certain elements tend to be retained over others, such as the wolf (or some other predator antagonist) and the deception and impersonation of the grandmother \citep{tehrani2013phylogeny}. These tend to be the core components, with important relations to most other parts of the story; removing these would cause a break-down of the story. More peripheral details that are not as intertwined in the narrative, such as the red hood itself or what is in the basket, are more easily lost. Different core components reinforce one another by together providing a coherent narrative. More peripheral details could potentially be maintained by for example grouping them together through rhymes. It has also been demonstrated that for example “minimally counterintuitive” elements (such as speaking wolves or pumpkin coaches) are more easily maintained \citep{boyer1994naturalness, norenzayan2006memory}. Quite possibly, the selective advantage of such properties could also be explained through a systems lens: concepts should be minimally counterintuitive, meaning that coherence is still key, and being counterintuitive sets them apart from the rest of the receiver’s belief system, yielding an element of surprise. They are distinctive but embedded, standing out enough while remaining inferentially connected to existing schemas, integrating with the narrative system instead of breaking it \citep{waddill1998distinctiveness, barrett2001spreading}.

A greater challenge is to describe what happens when the message itself is not at all replicated but triggers a reaction by association, such as the utterance “red is the colour of love” or pointing at a furry barking creature while repeatedly uttering the sound “dog” to a small child \citep{claidiere2014how}. In these cases, rather than the sender’s traits being replicated, concepts are being reconstructed by the receiver \citep{sperber1996explaining, miton2023cultural}. However, reconstruction must use information available to the receiver, that is, tokens the receiver already possesses in combination with the tokens exposed by the sender. Tokens from the sender activate related representations among the receiver’s tokens, thus there must be some relationship between those two sets of tokens. The social interaction involves not only selection among the sender’s traits, but also among those already in the mind of the receiver, specifically those most related to the cues passed on by the sender. The receiver selects a subset of activated and new traits, presumably those that fit well together and within the receiver’s ecosystem of beliefs. When incorporating new traits and recombining the modified cluster of traits, novel variants may emerge, with properties not necessarily represented in the constituent parts \citep[cf.][]{fauconnier2002way}. For examples, children build rule systems by hearing utterances, from which grammar emerges. Again, transformation might be viewed as a process of selective retention and recombination through a lens of cultural systems. This process does not only include surface tokens, such as a bowtie or the word dog, but also relational structures, for example when and with what apparel bowties are used, grammar, or the semantic range of “dog”.

\subsection{Dissonance}

How does the process work then when the mind selects, disposes or recombines traits? The human ability to store and process sequences of information provides basic prerequisites for organising beliefs. This organisation is needed to build functional models of the world \citep{johnson-laird1983mental}. Building a random structure in your head is quite useless (you cannot do anything with, say, catawampus + dongle = smurf) so the option would be simple associations and only learning direct responses to particular stimuli. The organisation of beliefs is needed for higher-order skills such as thinking and planning \citep{enquist2023human, heyes2018cognitive}. This in turn presumes that there is some mechanism for sorting information and filter out that which does not fit, to provide coherence \citep{thagard2000coherence}. There are many candidate psychological mechanisms that approximate such a mechanism. Some of these are balance theory, which holds that people strive for evaluative consistency in their relationships and attitudes to avoid imbalanced triads (if you like A and A likes B, then you are pushed to like B or otherwise also dislike A) \citep{heider1946attitudes, heider1958psychology}; motivated reasoning, where reasoning processes are unconsciously steered by desired conclusions, often leading individuals to accept congenial information and scrutinise uncongenial information more critically \citep{kunda1990case}; confirmation bias, the tendency to preferentially seek out, interpret and remember evidence that supports existing beliefs while downplaying contrary evidence \citep{nickerson1998confirmation}; and cognitive consonance and dissonance, where inconsistency among beliefs or between beliefs and behaviour produces psychological discomfort, motivating individuals to reject, change or reinterpret beliefs to restore consistency \citep{festinger1957theory}.

For example, an individual with a lifestyle demanding a large carbon footprint may be prone to seek out information that anthropogenic global warming is not an urgent issue, and still faced with evidence that it is, said individual may experience psychological discomfort that can only be sorted out by changing their lifestyle, find excuses for it, trivialising, or outright rejecting the evidence or its source \citep{simon1995trivialization, mcgrath2017dealing}.

These biases are not independent modules but overlapping manifestations of a more general coherence-seeking tendency. They can all be seen as different strategies to preserve internal consistency among beliefs and attitudes, and one bias might be inferable from another. Models of coherence as constraint satisfaction treat such biases as surface expressions of the same underlying process \citep{thagard2000coherence}. For example, cognitive dissonance provides motivation to resolve this lack of coherence, leading to motivated reasoning and confirmation bias to fulfil constraint satisfaction. The drive to avoid cognitive dissonance has ample empirical evidence \citep{elliot1994motivational, cooper2007cognitive} and it may be the process that is most reconcilable with a systems framework. This will thus be the term used in the remainder of this chapter. Dissonance can be seen as a form of system tension -- an imbalance among interconnected beliefs -- that drives the system towards relaxation by reconfiguring elements until coherence is restored. In this sense, cognitive dissonance reduction functions like a self-organising process, stabilising the belief system as a whole.

\subsection{Filters}

We build our worldviews, our own internal beliefs systems, incrementally. In contrast to genetic inheritance, cultural systems are not transmitted all in one go \citep{smolla2021underappreciated}. Rather, they are built gradually in a highly path-dependent process where each additional element can be evaluated and tailored to fit into the individual’s current ecosystem of ideas. When exposed to a new idea, whether that idea will make sense depends on how it relates to the individual’s other ideas. For example, a physicist trained in Newtonian mechanics may accept equations about force, mass and acceleration but not astrology as an explanation of motion, while someone with strong free-market convictions may welcome arguments for tax cuts but resist nationalisation. The other way round, introducing an additional god to explain a natural phenomenon may be reasonable to a polytheist but unpalatable to a monotheist, and universal healthcare fits within egalitarian worldviews of fairness and solidarity but less well within libertarian ones that emphasise individual responsibility and minimal state intervention. The processes of becoming a libertarian, egalitarian, theist, physicist or whatnot are in turn incremental ones based on some beliefs facilitating the acquisition of additional beliefs and inhibiting the acquisition of others.

Once you have acquired some beliefs, these constitute your lens to the world, influencing what other information to accept. One way to conceptualise this is to treat belief systems as networks of cultural traits, where links between traits have weights depending on their degree and direction of association. The links as such may be emergent and transmitted directly \citep{yeh2019cultural} or indirectly \citep{goldberg2018social}, or the beliefs themselves are transmitted, and to maintain logical consistency \citep{friedkin2016network} and/or reduce dissonance, this will result in updating individual beliefs and social connections \citep{galesic2021integrating} or more straightforwardly in acceptance or rejection \citep{jansson2021modelling}. Focusing on the transmission event, if new information invokes dissonance, then the simplest strategy is typically to reject that information. Previously acquired beliefs thus function as filters \citep{buskell2019systems, jansson2021modelling}. Information that passes the filter expands the individual belief system, with the consequence that the filter becomes more elaborate, as incoming information will have more pieces to fit.

In terms of cultural evolution, this idea gives flesh to how cultural selection operates: from the many variants encountered, those that fit the recipient’s current ecosystem of beliefs are preferentially retrieved, retained and re-expressed. From this, an effective fitness may emerge from pressures for coherence and inferential links. Ideas have high fitness to the extent that they are congruent with commonly propagated filters, that is, they cohere with other ideas in many local networks of beliefs. Accepted ideas reconfigure the filter itself by adding nodes and strengthening associations; this changes the local selection environment and makes further congruent ideas easier to accept \citep[a kind of niche construction; e.g.][]{kendal2011human}. Filters thus self-organise into path-dependent, increasingly discriminating structures that define attractor basins, that is, coherent packages towards which new adoptions (reconstructions) are pulled \citep[cf.][]{sperber1996explaining, morin2015how}.

\begin{figure}
    \centering
    \includegraphics[width=1\linewidth]{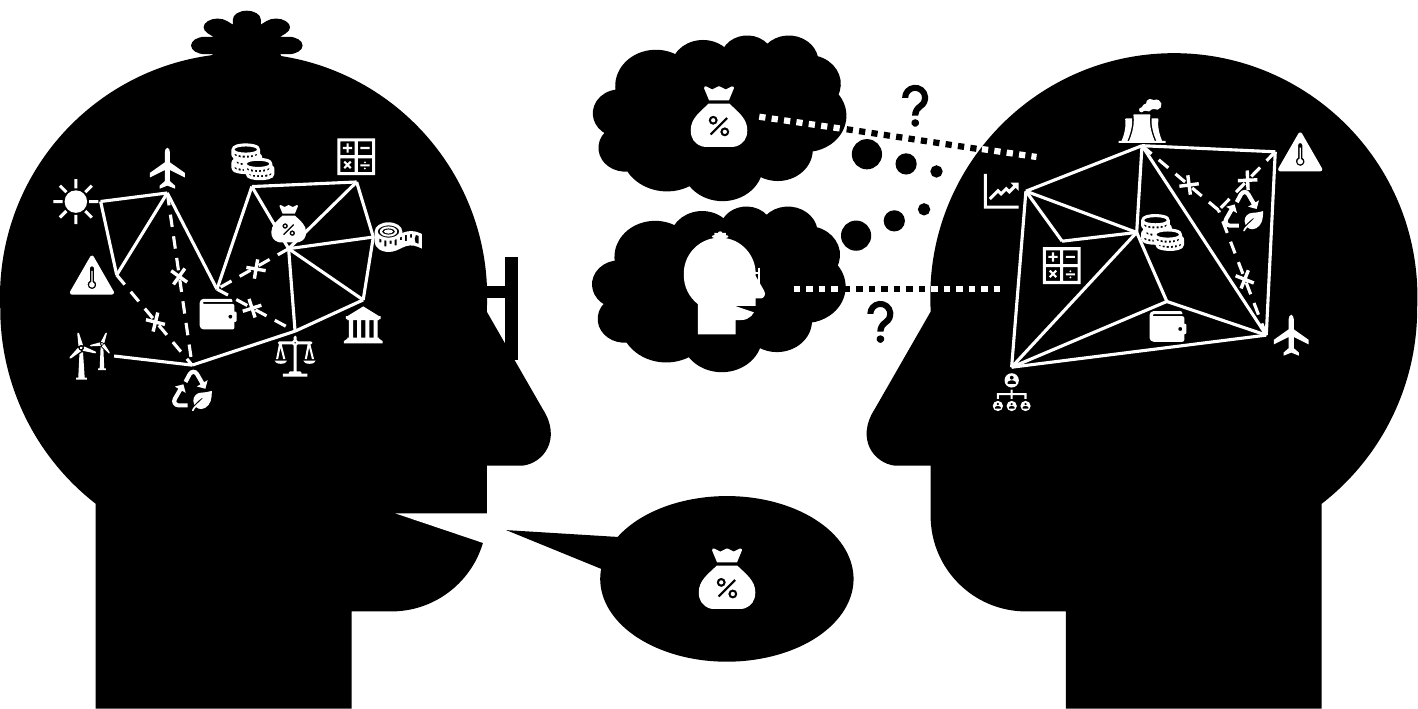}
    \caption{Information filtering. The sender exposes those beliefs most congruent with their self-image and its ideas about the receiver. The receiver accepts or rejects traits based on how they or the sender fit into their current belief system.}
    \label{fig}
\end{figure}

The evaluation of coherence or what creates dissonance should depend on the complexity of and familiarity with the topic, resulting in different types of filtering \citep{jansson2021modelling}. Some of these are illustrated in Figure \ref{fig}.

Some topics are highly accessible to the receiver, and they can evaluate or sense the coherence of new information based on its actual content, for example because it clashes with related experiences (drinking two cups of coffee before bed helps you sleep better; stealing from a friend strengthens your friendship), domain knowledge (a triangle can have two right angles; all mutations are harmful and decrease fitness; satellites need engines to keep moving; language must use word order to phrase questions) or simple logical consistency (2+2=5; an event happened before it occurred). We could thus label this process as content filtering \citep[c.f. direct bias;][]{boyd1985culture}.

Often, however, the validity of the information is hard to assess directly by content coherence, for example because the receiver lacks first-hand access to evidence or a sufficient specialist background. This may apply, for example, to reports on world events or even personal testimony of local events where the receiver was not around, or scientific theory and evidence. A reasonable heuristic is then instead to evaluate the credibility of the source, either a person or a media outlet, resulting in source filtering. There are many candidate processes for how this could operate, but one option is to evaluate the overall coherence of the perceived belief system of the sender with one’s own. If we seem to share core values and epistemological assumptions, or have similar needs and circumstances, then you are likely to have high quality information for me. There may also be social considerations here: sharing belief systems makes it easier to maintain social ties, and at the other end, deviating from the beliefs of your social circle may create social dissonance \citep{galesic2021integrating, dalege2025networks}. Either way, source filtering effectively turns the evaluation of information into an assessment of systemic coherence across individuals, making it a collective endeavour, giving rise to collective cognition.

Models of cultural evolution tend to focus on selection on the receiver side, but how the sender filters information determines what gets transmitted in the first place \citep{enquist2024cultural}. People have lots of information that they never pass forward, plausibly because this information inflicts some kind of dissonance, perhaps because it does not make sense or it violates the self-image the sender wants to convey. You may be familiar with the Flat Earth Society but find no reason whatsoever to pass on their ideas, or more happily talk about an avant-garde jazz concert you went to last month than your fondness of ABBA, just as a scientist might privately read horoscopes or an environmentalist might downplay their fondness for long-haul travel. This kind of filtering has been referred to as sender filtering \citep{jansson2021modelling}, but in symmetry with the receiver side, it could also be split into content-based and target-based filtering.

\subsection{Recombination and creativity}\label{sec:recombination}

These processes are not in principle restricted to social transmission but may apply also to individual learning. Exploration itself is subject to confirmation bias, filtering out evidence before it has a chance of being discovered, even in everyday life. In science, you see what your theory allows you to see \citep{hanson1958patterns}. Among the things you do explore, observations that fit into your belief system are easier to understand and incorporate, whether learning from interaction with the environment or individual motivated reasoning. Observations can also be reinterpreted to fit with your belief system \citep{kahan2017motivated, lord1979biased} or evaluated more harshly when they do not fit \citep{taber2006motivated}. Here are some potential everyday examples: a student who thinks they are ‘bad at maths’ may avoid mathematical tasks altogether, a gardener may explain a wind-bent plant as still reaching for the sun, or a music student may dismiss syncopation as a mistake because it violates their expectation that a bar resolves on the strong beat. Cognitively, the brain produces thousands of fleeting thought segments each day \citep{klinger2004motivational}, the vast majority of which never make it into conscious awareness, let alone long-term memory. As attention and working memory resources are limited \citep{baddeley2012working}, only a heavily filtered subset is retained and integrated into an individual’s belief system. Concepts survive when they connect to existing categories and support inferences \citep{campbell1960blind, boyer2001religion}; in this sense, the very emergence of ‘new’ cultural traits is already subject to systemic filtering long before any social transmission occurs. We could call this innovation filtering.

What is then an innovation? Creativity could be seen as part of the process of restructuring the belief system \citep{thagard2011aha}. This restructuring is continuously ongoing: we are exposed to ideas not only directly in social interaction and environmental exploration, but candidate ideas are generated from memory, analogy or blending and then selected or discarded by coherence-based filters \citep{campbell1960blind, thagard2000coherence}. Accepted combinations can rewire the system by changing the strength and direction of links and outfiltering elements that clash with the reconfigured whole to restore coherence or to simplify complex high-dimensional data into lower-dimensional representations with coherent concepts and categories \citep[ibid;][]{gardenfors2000conceptual}.

As elements are recombined or blended under compatibility constraints, sometimes new cultural traits can emerge, which, while being composed of preexisting entities, can have emergent properties not present in the individual parts themselves \citep{finke1992creative, fauconnier2002way}: for example, discovering a musical rhythm by tapping two pencils against a metal desk, mixing sweet and sour ingredients in cooking, playing with words and coining new metaphors or with chords and finding new harmonic progressions, “discovering” multiplication as repeated addition, and drawing analogies between viral spread in epidemiology and meme diffusion. These emergent ideas can then themselves become nodes that shape future filtering and attractor dynamics within cultural systems. At larger cultural scales, the same logic appears as combinatorial evolution of technologies and ideas \citep{arthur2009nature}.

Interestingly, this process has given rise to tools that speed up the creative process. It gave rise to techniques for storing information, such as memory techniques and later physical storage media, expanding the number of components available beyond what individuals can hold in working memory. This eventually led to mass media, from the printing press to the internet, enabling a faster exchange and further recombination of ideas. Tools for information processing, such as computers, further automated parts of this search, and today, generative AI can accelerate exploratory and combinational search without being explicitly instructed where to look and what to combine. In this sense, generative AI systems could be said to exhibit at least a limited analogous capacity to humans, by operating through recombination under compatibility constraints, sampling from learned structures to produce outputs that are coherent yet sometimes surprising \citep[e.g.][]{boden2004creative}. So far, however, this search is not open-ended, as current systems lack endogenous evaluation and restructuring. Humans still define the goals and perform the final selection, integration and system-level restructuring of model outputs \citep[e.g.][]{rainey2025could}. Then again, even human goals emerge through the same recursive processes of recombination and selection.

\subsection{Metafilters and superstructures}

As already hinted upon, some cultural filters have a more profound effect on acquisition, transmission and recombination than others. First, not all beliefs are created equal. Considering a package of beliefs, for example an ideology, as a cluster of connected nodes in a network, some nodes will have a higher degree of connections than others and perhaps occupy a central hub position: there are some core beliefs. Peripheral beliefs should be relatively easy to replace, since that would not require any major restructuring of the network but leave the belief package largely intact. Challenging a core belief, on the other hand, could have vast repercussions. If the belief is supported by its connections (i.e., it has a high indegree), then it would withstand most attempts, since challenging information would need to address those beliefs. If the core belief is rather characterised by supporting its connections (i.e. has a high outdegree) then challenging information should invoke severe cognitive dissonance if it cannot be easily rejected on its incoherence with other beliefs. Replacing a core belief may result in connected beliefs losing their support, which may then in turn also need to be revised, leading to an effortful major restructuring of the belief network.

Some core beliefs could be that there is a divine plan, hard work leads to success, people close to me have my best interests at heart, science and reason are reliable guides, widely held beliefs are true, truth is relative or hidden by powerful interests and my memories are accurate. In each case, challenging a core assumption triggers cascading revisions or persistent dissonance. A lifelong meat-eater who comes to see animal suffering as wrong, or a believer who doubts an afterlife, must revise not only specific views but their moral and existential framework. Likewise, successful scientific models such as heliocentrism or evolution incurred entire paradigm shifts.

Some traits can also have a direct effect on successive cultural transmission, on what can be acquired and at what rate, while they are themselves culturally transmitted. Such traits can be called regulatory traits \citep{acerbi2014regulatory}. Learning to read and write opens a whole new transmission medium. Preferences, learning abilities and openness are all properties that can be socially acquired and influence your receptivity to new traits. The same goes for teaching abilities and persuasiveness, which influence your effectiveness as a sender. In fact, such metaproperties suggest that traits that make you an effective sender drive transmission more than those making you an effective receiver, since the former operate during the very transmission event, in contrast to the latter \citep{enquist2024cultural}. If you have ever wondered why people are more keen to make themselves heard than to listen to what others have to say, then the dynamics of trait interactions may provide an explanation.

One type of traits that can regulate behaviour far into the future are goals. There are physiological reasons for why you want to find food, but why your search for saturation would involve spending hours on preparing sushi or a soufflé seems to do with culturally transmitted goals, emergent in cultural systems. The same goes for wanting to learn to play the piano, or to become a writer or an influencer, activities not necessarily helping you to spread your genes, but that may sit well in your cultural context. People learn what is worth pursuing from norms, social structures, narratives and tools, and these elements can combine into stable aims that function like regulatory traits. They act as internal drivers of information processing, directing attention and evaluation towards outcomes that confirm or advance existing aims, thus what we would identify as an expression of confirmation bias.

Many other traits can also influence the information processing and how to take trait relations into account. Cultural skills such as literacy, numerical aptitude, knowledge retention and dexterity, and critical thinking all have effects on what can be acquired later. Critical thinking and a belief in the hypothetico-deductive model emphasise logical coherence between traits, likely strengthening content filtering. Some belief packages are on the contrary supported by a decreased importance of consistency. Sanctifying unfalsifiable postulates such as “God works in mysterious ways” relaxes the requirement for coherence privileging other trait relations, stressing ritual stability, social cohesion and shared commitment to unquestionable premises that function as control mechanisms in symbolic systems \citep{rappaport1971sacred}.

When combined in large packages, regulatory traits can amount to fundamental human skills, so called cognitive gadgets. There is evidence that many of the cognitive mechanisms that distinguish humans, such as imitation, language, causal reasoning and theory of mind, are not biological givens but culturally acquired ``gadgets'' (\citealp{heyes2018cognitive}; see \citealp{jansson2026cultural} for a summary). These skills themselves are cultural products, spread through social learning, and then strongly influence how subsequent information is processed and transmitted, with a drastic impact on cultural evolution. They thus function as a kind of metafilters, regulating both perceived relations between traits and the type of filtering involved. For example, causal reasoning and planning stresses content relations, whereas overimitation and culturally scaffolded mindreading could sharpen source filtering by focusing on social affiliation, in the first case \citep{lyons2007hidden}, and intentions and reliability cues, in the second case (\citealp{heyes2018cognitive}; cf. \citealp{sperber2010epistemic}, on epistemic vigilance).

When interacting traits reinforce one another and give rise to stable new functions, they can form higher-order nodes -- superstructures that regulate large parts of the cultural system. Cognitive gadgets are clear examples. Skills such as language, imitation, causal reasoning and theory of mind emerge from the coordination of more basic capacities, including memory, perception, motor control, attention and social motivation, and they in turn exert a strong influence on later learning and communication. Once established, such higher-order nodes shape how new information is filtered and combined: language structures thought and transmission, causal reasoning organises inference and planning, and shared intentionality underpins norm enforcement and coordination. Comparable processes occur when norm systems or institutions become stable clusters that define what kinds of beliefs and behaviours are acceptable within a group. The structure of such complexes can be formalised with hypergraphs or higher-order networks, which are networks where a link can be drawn to a whole collection of nodes, thus capturing how sets of traits jointly constrain and regulate other nodes in the network \citep{battiston2020networks}.

Just as traits can be assembled into higher-order nodes, with the same principles, it should, at least in theory, be possible to disentangle them into clusters of lower-order nodes. A chord could be seen as a trait and analysed in terms of its function in a harmonic progression, which is in turn part of a tune. At the same time, a chord is composed of tones, which in turn has a timbre given by the overtone series, combining frequencies with different strengths. Words are combined into sentences and stories, and are composed of syllables and phonemes. This may clarify how to relate to one of the million dollar questions in cultural evolution: what is a cultural unit? It is a cluster of interrelated traits. This of course defines a somewhat impractical infinite recursion, with turtles all the way down. A pragmatic response could be that you decide on the level to place the cultural unit depending on the application.

\section{Emergence}

We have already seen several examples of how the combination of cultural traits can lead to emergent properties that the traits do not have on their own, for example when they form regulating metafilters, make belief systems stable or incur a major restructuring of a belief network. Many of these processes would be difficult to study endogenously, without making strong assumptions on biases or environmental preconditions, unless we take a systems perspective. When considering trait relations and their interactions through filtering, however, many cognitive and social phenomena can be grown endogenously in models with minimal cognitive prerequisites along with purely cultural evolution.

\subsection{Complex patterns}

Culture is a complex adaptive system \citep{miller2009complex} and studying it as such means that we can also trace the generative processes of the patterns that complex systems exhibit, patterns we do observe both in the mind and society. There is a chapter of this volume that provides an introduction to the relevant concepts \citep{michaud2026complex} and there are more in-depth treatments (see \citealp{thurner2018introduction} for a mathematical treatment; \citealp{ladyman2020what} for a less technical one). We will thus only touch upon a few illustrative examples.

In general, a complex system is a system composed of many components that interact with one another. Such systems give rise to emergence, nonlinearity, path dependence, self-organi\-sation and nested structures, among other properties. The previous section gave examples of higher-order traits, with emergent properties and giving rise to nested structures with hierarchies. We have also seen examples of the other properties in previous sections.

Nonlinearity arises because small, local adjustments can tip densely connected clusters into new configurations once thresholds are crossed. A single change in how one belief or practice fits with others can spread through the system as adjustments ripple across the network. A small drift in pronunciation of a vowel can spread through an entire vowel inventory, as in the Great Vowel Shift in Middle English (as presented above in Section \ref{sec:embeddings}), and small cracks in a scientific worldview can accumulate until a full paradigm overturns, as when heliocentrism replaced the Ptolemaic system or germ theory replaced miasmas. Small changes can lead to tipping points after which positive feedback loops amplify change, such as when people stay silent until enough others speak up: during the Arab Spring or the \#MeToo movement, a public act of dissent or testimony may encouraged the boldest, and a quiet background of hesitation turned into a wave of participation when people felt safe enough \citep{granovetter1978thresholda, howard2011opening, gonzalez-bailon2011dynamics}. The counterpart is negative feedback, which stabilises systems, for instance when information filters protect belief packages by rejecting incompatible information, or when norms, peer review, or community moderation push behaviour back towards accepted standards, such as regimes responding fiercely to public acts of dissent.

Path dependence arises because early conditions shape what comes later. Standards such as rail gauges, keyboard layouts and measurement systems lock in whole ecosystems of practices. Gradually evolving standards such as grammar and writing systems also restrict later options, and the same logic applies to individual acquisition of beliefs and skills. Once a belief system or skill set has begun to take shape, new elements are filtered through what is already established, and those that do not fit are less likely to take hold. Beliefs cluster in mutually compatible packages. Early acquired beliefs thus build filters that direct us towards certain belief packages rather than others, and early training guides which abilities we continue to invest in and refine. What we learn first sets the stage for what we can easily learn next and potentially closes the door to other paths.

Self-organisation occurs when order emerges from the mutual adjustment of many interconnected traits. As beliefs, practices, and norms are filtered through one another, coherence grows locally until larger-scale patterns appear, without any central design. Languages, moral codes, or scientific paradigms thus stabilise because their internal relations support one another \citep{deboer2005selforganisation, anzola2017selforganization}: changes that fit with surrounding traits are retained, others fade out. The same logic applies in social movements, where each local act of expression or coordination strengthens the credibility of similar acts elsewhere. When enough local alignments accumulate, a collective pattern, such as shared norms, common goals, or a unified voice, takes shape on its own.

These properties are important when trying to explain and predict cultural change. If dynamics are nonlinear and path dependent, then the same intervention (a new practice, a policy, a norm) will have different effects depending on how it fits and where it lands in the cultural system: near a threshold it may trigger a cascade; far from one it may be neutralised by surrounding constraints. And if much of the observed order is self-organised, then, rather than smooth, incremental trajectories, we should expect long periods of stability while local relations maintain coherence, interrupted by rapid cascades when accumulated system dissonance forces large-scale restructuring.

The patterns are hard to generate in conventional single-trait models. Treating traits as independent with fixed transmission biases typically cannot (i) generate threshold effects and phase transitions, because of the lack of feedback between trait--trait relations, (ii) reproduce historical lock-ins and path dependence, because the state of one trait does not reshape the selective environment for others, or (iii) yield endogenous organisation of higher-order packages, because there is no mechanism for sets of traits to mutually stabilise or destabilise one another.

\subsection{Cultural niches}

Emergent properties of complex cultural systems lead culture to reorganise its own evolution. This occurs also in biological ecosystems when organisms alter their local environment (e.g. beavers building dams or earthworms physically and chemically modifying the soil), which can then feed back to the selection pressures on a recipient organism, often the organism itself. This process is called niche construction \citep{odling-smee2003niche, odling-smee2024niche}. Humans are of course a prime example, in how we have transformed the living conditions on the entire planet. This has modified genetic selection through gene-cultural co-evolution \citep{laland2010how}, for example for lactase persistence through farming \citep{malmstrom2010high}, but cultural systems can also build their own niches.

Some examples are norm systems and the path dependent lock-ins in technological standards mentioned above. Actors in the social world can even make use of this, whether deliberately or incidental. Technological ecosystems such as Android and Iphone, or different online social networks, are prime present-day examples. Once a platform gains traction, its apps, device compatibility, accessories and connected user base create a self-reinforcing niche that locks users into the system and consolidates its position among them. Other examples are economic and normative systems which construct institutional and moral niches that stabilise particular practices. They do so not only at the level of material and institutional structures, but also in how information is organised and transmitted.

Epistemic systems such as scientific paradigms illustrate this clearly: they stabilise concepts, methods, and standards that determine which ideas can meaningfully develop \citep{kuhn1977objectivity, benedetto2024theory}. In a similar way, cultural systems construct cognitive environments that guide how individuals learn and think. During cognitive development, this occurs through the structuring of children’s learning environments via pedagogy, artefacts, routines and social practices \citep{flynn2013target}, equipping them with cognitive gadgets that further reshape both the physical and informational environment.

Belief packages likewise reshape their informational and developmental niches. Through behaviours that restructure interaction partners (e.g., through homophilous ties), information channels and evaluation standards, they alter what information is encountered, trusted and practised next \citep{creanza2014complexity}. These constructed niches bias attention and interpretation, modifying our filters so that some ideas become increasingly easy to accept while others are excluded, producing path dependence and attractor basins in belief space. In this sense, belief systems externalise constraints that then internalise as cognitive tools and regulatory filters; in what could be described as epistemic niche construction \citep{sterelny2010minds, benedetto2024theory}, accepted beliefs reconfigure the very selection environment (through selective exposure, epistemic trust, and coherence restraints) that the next beliefs must traverse. In this way, our minds can be attuned to, for example, conservative, liberal or socialist values.

When such processes unfold collectively, they can partition the population into relatively self-contained domains of meaning. Beliefs that fit well together reinforce each other and the informational environments that sustain them, leading to clustered belief systems that stabilise distinct epistemic niches. Through information filtering, initial differences in belief can push individuals towards different informational niches that become increasingly resistant to external input. Once established, these niches can become echo chambers \citep{nguyen2020echo, cinelli2021echo}, which are socially and cognitively constructed environments where a person only encounters information or opinions that reflect and reinforce their own.

The term ‘echo systems’ is often used in the context of news and social media. In the US, for example, the Fox News centred ecosystem functions as a relatively insular niche that privileges congenial sources and discounts mainstream outlets \citep{benkler2018network}. Kremlin-linked outlets such as RT similarly frame external sources as untrustworthy, drawing their audience into a shared narrative space through identity-consistent messages \citep{soares2023falling, shirikov2024rethinking}. In practice, media diets function as epistemic niches: they stabilise source and content filters through repeated, routinised exposure to congenial outlets, making cross-cluster uptake rarer and locking belief change into local attractors. Social media further allows for homophilous ties to form, where the users not only select into informational niches, but also find interaction partners based on shared opinions \citep{cinelli2021echo}. Modern forms of algorithmic curation intensify these dynamics by automating selective exposure and trust calibration, thereby increasing partisan exposure \citep{dejean2022partisan} and reproducing and accelerating processes in manual information filtering \citep{santos2021link, bellina2023effecta}.

\subsection{Groups}\label{sec:groups}

The clustering of people with similar beliefs provides a mechanism of cultural group formation, where groups become demarcated through information exchange, as products of polarisation. If beliefs are intrinsically connected such that they form clusters and people filter information based on coherence, then the population can easily polarise along those clusters (\citealp{jansson2021modelling}; see also \citealp{friedkin2016network, galesic2021integrating, dalege2025networks}). This works through the sensitivity of initial conditions coupled with path dependence, that is, early acquisition of beliefs can steer people into particular ecosystems of beliefs, and even if people are later exposed to similar information, beliefs can further diverge \citep{nielsen2021persistent}. For example, ideologies can be internally mostly consistent, but which one you end up adopting may depend on early beliefs concerning human nature, social forces and fairness. The same applies to scientific paradigms or for that matter epistemic trust in scientific authorities at all versus conspiracies.

However, many beliefs are not intrinsically related but still tend to co-occur in polarised groups. Being concerned about climate change and Covid-19 are associated beyond ideology and scientific trust \citep{latkin2022association} and partisanship is associated with a wide range of beliefs and attitudes, such as support for particular energy sources \citep{brugidou2023return, isaksson2024political}, factual circumstances such as whether climate change is anthropogenic and the direction of trends in violent crime \citep{kahan2016politically}, well, even with drinking latte. Beliefs associations can also emerge through cultural transmission. If people employ source filtering, that is, estimate information based on the perceived credibility of the source rather than the content itself, then associations can easily emerge among initially unrelated beliefs and people channel themselves into disjoint belief packages, as in the case of intrinsic relations (\citealp{jansson2025emergence}; see also \citealp{goldberg2018social, galesic2021integrating}). This is also coupled with emergent group signals: someone who believes P can also be assumed to believe Q and R, and thus be part of a specific epistemically defined group, while someone who disbelieves P can also be assumed to disbelieve Q and R, not because of logical or other external constraints, but based on cultural associations. Group signals can in turn be a breeding ground for group discrimination \citep{jansson2015what}.

On the other hand, if there are strong relations between beliefs across clusters, then that would suggest that common understandings are possible. Let us conclude this section with an empirical example of this. Perhaps surprisingly, given the prevalent political polarisation, public opinion in the US has moved over time towards more liberal and progressive stances on moral issues among both liberals and conservatives \citep{mulligan2013dynamics}, for example in accepting gay marriage and opposing corporal punishment. It turns out that the same stances can be supported by individualising arguments based on care and fairness \citep{strimling2019connection}, core values for both liberals and conservatives \citep{graham2009liberals}. Meanwhile, traditionally conservative stances are mainly supported by group binding arguments based on authority, loyalty and purity, core values only for conservatives. Given the dynamics of belief systems and filtering described in this chapter, we might then expect arguments for conservative stances to be outfiltered by liberals, while arguments for progressive stances are easily incorporated in the liberal belief systems and induce cognitive dissonance for conservatives, which may eventually lead to acceptance. In line with this, it has been found that only conservatives are susceptible to group binding arguments, while both groups are indeed susceptible to individualising arguments \citep{jansson2025susceptibility} and that the strength of the connection between moral stances and their support by individualising arguments predicts public opinion change \citep{strimling2019connection}.

\subsection{Norms and institutions}

We have already, on several occasions, treated the relationship between trait interactions and higher-level norms that regulate those, and there is a full chapter dedicated to norm dynamics in this volume \citep{jansson2026emergence}. However, it is worth highlighting social norms as emergent products in cultural systems. Social norms can be viewed as emergent regularities \citep{hawkins2019emergence, przepiorka2022how} that crystallise when interdependent beliefs, practices, and artefacts mutually stabilise through everyday coordination and filtering. Because they align expectations across individuals, they are also the mechanisms that make groups operate as coordinated units, by regulating contribution, deference and sanction \citep{ostrom1990governing, bicchieri2006grammar}. Local judgements of “what fits” (information filtering) aggregate into shared expectations about “what one does”, from dress codes becoming legible signals within particular cultural packages to peer review conventions in science; once established, such expectations guide subsequent uptake and are reinforced by path dependence and threshold dynamics.

Seen through the analytical lens (and filters, to use a metareference) of this chapter, in line with structuralist perspectives emphasising relational organisation and shared meaning \citep{durkheim1912elementary, levi-strauss1963structural} and with later sociological elaborations on how symbolic order and social fields distribute authority \citep{bourdieu1977outline, foucault1977discipline}, norms thus function as meaning-making (they tell us how to interpret signals, roles, and reasons) and power-distributing (they allocate credibility, authorise speakers, and define legitimate moves within a practice) devices. They shape who is heard and how claims must be framed to be receivable, often via the informational niches they co-produce, such as media diets, platforms, and affiliations that amplify congenial sources and discount others, thereby structuring collective cognition and influence.

When codified and scaffolded, these normative patterns become institutions \citep{ostrom1990governing}, that is, formalised rule systems, procedures, and infrastructures (schools, journals, courts, platform policies) that function as population-level regulatory traits. Institutions enforce constraints (through monitoring, sanctions, credentialing etc.), redesign informational pathways (through creating information channels, curricula, peer review, recommendation systems etc.), and so feed back on cultural evolution by reshaping filters, stabilising some trajectories and making others costly or unthinkable. To once again refer back to multilevel interactions, institutions are themselves emergent products of cultural systems that, once in place, restructure the selection environment for future beliefs and practices, locking in standards or enabling rapid reorganisation when systemic tensions make piecemeal patching untenable.

\subsection{Artificial intelligence}

Language is arguably a complex adaptive system with grammar and meaning emerging from patterned relations among words and between users \citep{beckner2009language, kirby2015compression}, encoding cultural knowledge and practices \citep{tomasello2010origins, christiansen2016creating}. Generative AI is an emergent technology, built from countless technological and other cultural components that has made use of the information encoded in language. Systemic properties of language and culture make modern large language models possible \citep{tenenbaum2006theorybased, berti2025emergent}. Trained by next-token prediction on large text corpora, they perform a system-level compression of linguistic regularities, developing an internal embedding space in which distances and directions reflect associations between concepts through their recurring co-occurrences and contrasts \citep{mikolov2013efficient}. The result is not a dictionary of definitions, but a form of semantic competence, with a capacity to reproduce the structured associations among words and contexts that, in human communication, allow speakers to coordinate shared meanings. When the model predicts a token, it is subject to constraint satisfaction within this relational space: candidate continuations are accepted when they cohere with distributed constraints set by syntax, style, topic and discourse. Alignment procedures, safety layers and decoding rules \citep{ouyang2022training} act as filters that steer which regions of the space are routinely explored. LLMs are both products of cultural evolution and new agents that operate through the same systemic properties of language that culture has sedimented over time.

AI technologies can have a tremendous effect on cultural evolution through how they take over information filtering. Humans have outsourced filtering processes to technology before, for example by carrying out and verifying computations through calculators. Such technologies, however, are deterministic. The same input always gives the same output, and the way it does so is through well-known and widely understood algorithms, monitored by designers and users all around the world. We can trust the output of the calculator to the extent that we trust the sensibility of some specific rules or that malfunctioning would become known. However, recommender algorithms and generative AI automate filtering beyond such control.

Recommender algorithms select content for us through selective exposure. Early algorithms were based on feature similarity between items, a kind of content(-based) filtering, as described in this chapter, and also the term for such algorithms. However, collaborative filtering took off in the late 90s and has become more widely used \citep{gheewala2025indepth}. Instead of comparing items, it compares users and make recommendations based on the preferences of similar users. Behind the scenes, this is thus an automated form of source filtering where algorithms determine whom to trust. As discussed in Section \ref{sec:groups}, this can promote polarisation and lead to the emergence of cultural trait associations \citep{jansson2025emergence}.

Generative AI takes automated filtering to a whole new level, coupled with automated content creation. It remains an open question how they will influence cultural diversity: Through strong path dependence, LLMs could conserve norms, values, biases and other present-day cultural structures, as this is what they are trained on and then reproduce globally, also shaping the data for the training of future LLMs within these structures. However, as discussed in Section \ref{sec:recombination}, new viable recombinations may emerge, leading to substantial innovation. Whether innovative or not, more and more content is produced by AI \citep[e.g.][]{liang2025quantifying, spennemann2025delving}, overwhelming us with an abundance of information that increasingly requires automated filtering to be navigated. On the producer side, we may also be required to increasingly use LLMs to filter our thoughts into comprehensive texts that can cut through the noise. The automation of filtering also opens the opportunity for more localised and even personalised filtering adapted to small groups of people (compared to, e.g., traditional mass media), which could lead to fragmentation into small epistemic niches. In a feedback loop, all the fundaments of cultural evolution -- creation, transmission and selection -- may be increasing mediated by AI as we become more dependent on it \citep[cf.][]{brinkmann2023machine, kulveit2025gradual}.

\section{Conclusions}

This chapter concludes a book seeking a minimal account of cultural evolution: from the origin of uniquely human capacities to the drivers of societal change. A few minimal ingredients may suffice. Biological evolution endowed humans with the minimal prerequisites for culture: robust and precise sequence representation and domain-general learning abilities \citep{vinken2026similarities,lind2026age}; these support open-ended combinatorial representation and, with language, enable precise, incremental transmission of complex skills and ideas \citep{jon-and2026evolution}. This combination is the keystone of human cultural evolution and potentially explains its rarity in other species \citep{jansson2026cultural,vinken2026similarities,lind2026age,apel2026evolution,jon-and2026evolution}.

Once traits could be precisely transmitted and recombined, they assembled into interdependent systems whose internal relations confer meaning and constraint \citep{jon-and2026evolution}. Cultural traits do not travel as isolated units: they cluster into co-evolving multilayer networks of beliefs and practices. In such systems, previously acquired traits and infrastructures become filters that introduce coherence constraints and bias attention, trust, memory, and production. This endogenous filtering, whether directly based on content, indirectly through source filtering on the receiver side, or target filtering on the production side, accumulates local coherence pressures into population-level selection: variants that fit more belief ecosystems are preferentially retained and re-expressed. Over time, these filters self‑organise into cultural attractors \citep{jansson2026cultural} and epistemic niches that both stabilise cultural packages and, when tensions arise through dissonant beliefs or new recombinations, enable rapid reconfiguration, when small local adjustments can trigger large‑scale cascades \citep{michaud2026complex}.

As filters evolve, they give rise to fragmentation through emergent properties such as homophily and selective exposure that shape media diets, platforms, and affiliations, reinforcing clustering and paving the way for group formation. At the group level, norms crystallise as emergent regularities that align expectations and regulate sanctioning, forming group-level traits in which filtering becomes a collective endeavour embedded in shared norms and coordinated behaviour \citep{jansson2026emergence}. Institutionalisation then codifies these normative filters into broader societal structures of communication and governance: curricula, laws, schools, languages, mass media, governments, and other infrastructures that collectively reshape information pathways and coordination mechanisms. For example, fertility norms reshape population structure and transmission \citep{kolk2026population}.

Seen through the cultural systems framework, this general pattern of interaction shares one underlying logic: combinatorial creation, systemic coupling, and endogenous filtering generate emergent order \citep{michaud2026complex}. The framework reconciles selection and transformation by locating selection pressures inside culture, and it reframes transmission biases as products of prior cultural acquisition, with culturally evolved cognitive gadgets as intermediaries, rather than innate heuristics \citep{jansson2026cultural,jon-and2026evolution}. It also integrates structuralist attention to patterned relations with evolutionary attention to variation and retention, making the study of related phenomena tractable through formal tools, most notably mathematical modelling, which allows us to conceptualise, formalise, and investigate the emergent patterns that such systems produce, and probing thresholds and equilibria that verbal reasoning alone leaves opaque \citep{jansson2026modelling}.

Finally, human cultural systems have produced semi-autonomously self-regulating artificial ones that co-evolve with manual processing of trait relations. Recommender systems automate source filtering at scale, and generative models compress and recombine cultural structure to produce new content; both now participate in our transmission environment, accelerating horizontal transmission \citep{kolk2026population} and increasingly mediating creation, exposure and selection. As human and artificial systems co‑evolve, they form a joint cultural machinery in which cognitive, social and algorithmic processes become mutually dependent components, filtering and recombining trait relations into increasingly integrated adaptive structures. AI might become a game-changing actor in cultural evolution, amplifying and reshaping the very filters and recombination processes that cultural systems rely on.

\bibliographystyle{fredrik}
\bibliography{inpress,references}

\end{document}